\documentclass[journal]{IEEEtran}
\usepackage{color}
\usepackage{multirow}
\usepackage{cite}
\usepackage{balance}
\usepackage{amsmath}
\usepackage{enumitem}
\usepackage{amsfonts}
\DeclareMathAlphabet{\mathpzc}{OT1}{pzc}{m}{it}
\setlist[itemize]{noitemsep, topsep=0pt}
\usepackage{wrapfig}
\usepackage{pdfsync}
\ifCLASSINFOpdf
\usepackage[pdftex]{graphicx}
\DeclareGraphicsExtensions{.pdf,.jpeg,.png,.jpg}
\else
\usepackage[dvips]{graphicx}

\DeclareGraphicsExtensions{.eps}
\fi
\graphicspath{{figs/}}
\usepackage{epstopdf}
\usepackage{booktabs}

\usepackage{amsmath}

\begin{document}
	
	\title{Critical Load Restoration using Distributed Energy Resources for Resilient Power Distribution System}

	\author{Shiva Poudel,~\IEEEmembership{Student Member,~IEEE,}
		and Anamika Dubey,~\IEEEmembership{Member,~IEEE,}% <-this % stops a space
		\thanks{S. Poudel and A. Dubey  are with the School of Electrical Engineering and Computer Science, Washington State University, Pullman, WA, 99164 e-mail: shiva.poudel@wsu.edu, anamika.dubey@wsu.edu.}% <-this % stops a space
	}

	\maketitle
	
	%\mathrm{}
	\vspace{-0.3 cm}
	\begin{abstract}
		Extreme weather events have a significant impact on the aging and outdated power distribution infrastructures. These high-impact low-probability (HILP) events often result in extended outages and loss of critical services, thus, severely affecting customers' safety. This calls for the need to ensure resilience in distribution networks by quickly restoring the critical services during a disaster. This paper presents an advanced feeder restoration method to restore critical loads using distributed energy resources (DERs). A resilient restoration approach is proposed that jointly maximizes the amount of restored critical loads and optimizes the restoration times by optimally allocating grid’s available DER resources. The restoration problem is modeled as a mixed-integer linear program with the objective of maximizing the resilience to post-restoration failures while simultaneously satisfying grid’s critical connectivity and operational constraints and ensuring a radial operation for a given open-loop feeder configuration. Simulations are performed to demonstrate the effectiveness of the proposed approach using IEEE 123-node feeder with 5 DERs supplying 11 critical loads and IEEE 906-bus feeder with 3 DERs supplying 17 critical loads. The impacts of DER availability and fuel reserve on restored networks are assessed and it is shown that the proposed approach is successfully able to restore a maximum number of critical loads using available DERs.
		%Withstanding and recovering from the high-impact low-probability (HILP) events is a challenging problem for the aging and outdated power distribution systems.
	\end{abstract}
	

	\IEEEpeerreviewmaketitle
	\vspace{-0.2 cm}
	\section*{Nomenclature}
	\addcontentsline{toc}{section}{Nomenclature}
	\begin{IEEEdescription}[\IEEEusemathlabelsep\IEEEsetlabelwidth{$V_1,V_2,V_3$}]
		%\item [Acronyms]
		%\item[DER] Distributed Energy Resources
		%\item[CL] Critical Load
		%\item[RSN] Restored Sub-tree Network\\
		%\item[MILP] Mixed Integer Linear Programming\\
		\item[$G =(V,E)$] Graph representing the distribution system
		\item [$V$] Set of nodes in $G$
		\item [$E$] Set of edges in $G$
		\item[$C_l$] Set of nodes containing critical loads
		\item[$M$] Set of nodes with DERs
		\item[$Y$] Set of children nodes of node $i$
		\item[$L_p$] Set of nodes that belong to a loop/mesh in $G$
		\item [$\mathcal{P_\alpha}$] Set of parent nodes of node $i$ along path $\alpha$
		\item[$n(.)$] Cardinality of a set
		\item[$\alpha$] Index of ossible paths in a loop/mesh
		\item[$v_i^k$] Node-DER assignment variable
		\item[$y_{i,\mathcal{P_\alpha}}^k$] Node-path assignment variable
		\item[$s_i$] Critical load pick-up variable
		\item[$R_p$] Restoration path reliability
		\item[$S_k$] $k^{th}$ Restored Sub-tree Network (RSN)
		
		\item[$a_{DER}^k$] Availability of $k^{th}$ DER
		\item[$T_k$] Restoration time for $k^{th}$DER
		\item[$E_k$] Reserve energy of $k^{th}$ DER at the time of outage in kWh
		\item[$l_k$] Number of distribution lines in $k^{th}$ RSN
		\item[$n_{\mathcal{P_\alpha}}$] Number of parent nodes along path $\alpha$
		
		\item[$U_R$] Effective restoration unavailability
		\item[$r_{ij}$] Resistance of branch joining nodes $i$ and $j$
		\item[$x_{ij}$] Reactance of branch joining nodes $i$ and $j$
		\item[$P_i$] Active power demand at node $i$
		\item[$Q_i$] Reactive power demand at node $i$	
		\item[$p_{i,\alpha}^k$] Injected real power at node $i$ via path $\alpha$
		\item[$q_{i,\alpha}^k$] Injected reactive power at node $i$ via path $\alpha$
		\item[$V_{i,\alpha}^k$] Voltage of node $i$ along path $\alpha$
		\item[$p_i^k$] Active power injected into node $i$ in RSN $S_k$
		\item[$q_i^k$] Reactive power injected into node $i$ in RSN $S_k$
		\item[$V_i^k$] Voltage of node $i$ in RSN $S_k$
		
		\item[$P_{max}^k$] Active power capacity of $k^{th}$ DER in kW
		\item[$Q_{max}^k$] Reactive power capacity of $k^{th}$ DER in kVar
	\end{IEEEdescription}
	\vspace{-0.2 cm}

\section{Introduction}
	
	\IEEEPARstart{E}{lectricity} networks are one of the most critical infrastructures of a nation. Unfortunately, the extreme weather events can disrupt the electricity supply for an extended time resulting in the loss of critical services for days and sometimes even for weeks and severely affect the customer safety and security. According to \cite{campbell2012weather} approximately 78\% of the outages from 1992 to 2010 were caused by extreme weather events affecting around 178 million metered customers and costing US economy 18 to 33 billion dollars per year \cite{salman2015evaluating}. Moreover, 90\% of the customers were affected due to the damages in the power distribution feeders. This calls for the critical need to improve distribution system resilience for natural disasters.
	
	Grid resilience is characterized by its ability to withstand and recover from the high-impact low-probability events \cite{panteli2015grid}. One of the requirements for a resilient distribution system is the ability to restore power to the critical loads for the duration of the outage. During a natural disaster, when the main power grid supplying distribution system is unavailable, the traditional distribution system restoration approaches are inapplicable \cite{solanki2007multi, lim2006service, nguyen2012agent}, calling for advanced system restoration methods. One such approach is to utilize distributed energy resources (DERs) as community resources by extending their zone of service to other loads.
	
	In literature, multiple articles have sought to improve grid resilience by using microgrids and DERs \cite{dong2014investigation, arghandeh2016definition, gaber2016heuristics, mohagheghi2011applications, li2014distribution, gao2016resilience, chen2016resilient, wang2015self, farzin2016enhancing, pesgm2017}. Methods are proposed to improve service reliability by isolating microgrids during faults to serve the local loads \cite{dong2014investigation, arghandeh2016definition,gaber2016heuristics}. Researchers have also sought to improve the grid resilience by using microgrids to restore not only local loads but also the critical loads in the distribution feeders \cite{ mohagheghi2011applications, li2014distribution, gao2016resilience,chen2016resilient, wang2015self}. For example, a restoration algorithm based on spanning tree is proposed to restore the critical loads using microgrids and maximize the duration of the restored loads \cite{li2014distribution},\cite{gao2016resilience}. In \cite{chen2016resilient} microgrids energized by DERs are formed to restore loads after a major outage. In \cite{wang2015self}, the healthy portion of the distribution system is sectionalized into self-sustained microgrids to continuously provide the power supply to a maximum number of customers. A two-level hierarchical outage management scheme is developed in \cite{farzin2016enhancing} for the resilient operation of multi-microgrids without restricting their autonomy.
	
	The aforementioned literature, however, has one or more limitations. A majority of literature fail to include the impacts of potential failures within the distribution system after the restoration plan has been executed \cite{dong2014investigation, gaber2016heuristics, arghandeh2016definition, mohagheghi2011applications, li2014distribution, gao2016resilience, chen2016resilient, wang2015self, farzin2016enhancing}. The distribution feeders are more likely to fail in the aftermath of a natural disaster. The grid resilience realized through DER enabled restoration not only depends upon the capacity and duration of the restored critical loads but also upon the robustness of the restoration plan to post-restoration failures. In \cite{pesgm2017}, we proposed a mixed-integer linear program (MILP) to maximize the restoration availability of the critical loads by forming self-sustained islands that are robust to post-restoration failures. The prior work, however, was limited in its consideration of several critical topological and operational characteristics of an actual distribution system. The approach presented in this paper provides a more generic framework for restoring critical loads using DERs in the aftermath of a natural disaster. The specific contributions of this paper are further stated in Section I.A. Furthermore, most of the proposed restoration methods are based on path search and heuristic approach \cite{gaber2016heuristics, mohagheghi2011applications, li2014distribution, gao2016resilience}. Other methods, do not consider the possibility of tie-switches or alternate paths serving loads in the formulation \cite{chen2016resilient}, \cite{wang2015self}, \cite{pesgm2017} and few do not consider the duration of critical load restoration in the optimization formulation \cite{chen2016resilient},\cite{wang2015self}, \cite{pesgm2017}.
	
	In this article, we propose a framework to restore power supply to the critical loads in an event of a major disaster by optimally utilizing DERs. A resilience metric is defined that includes the impacts of post-restoration failures in distribution lines and DERs. The proposed MILP framework restores critical loads in the feeder while: 1) maximizing the post-restoration reliability of restored loads, 2) including tie-switches and open-loop distribution system configurations into the optimization formulation, and 3) optimally allocating DERs for an equitable restoration of the critical loads.  
The proposed framework is applicable to all dispatchable DERs with specified capacity and reserve energy. DER capacity is defined as the rated real ($P_{max}$) and reactive power capacity ($Q_{max}$) of DER. DER reserve energy ($E_k$) indicates the reserve energy in kWh available for each DER immediately after a disaster. Note that the specific concerns related to the dynamics of different types of DERs during an islanded operation is beyond the scope of this work.

	\vspace{-0.3cm}
\subsection{Contributions}
The major contributions of this paper are as follows:	
\begin{enumerate} [noitemsep,topsep=0pt,leftmargin=*]
		\item \textit{Resilience for Post-Restoration Failure} - The restoration objective is defined as a function of post-restoration failure. The optimization problem maximizes the resilience of the restored networks by decreasing the likelihood of post-restoration failures. The proposed approach picks up a maximum number of critical loads during an arbitrary disaster condition depending upon DER capacity and available restoration paths.
		\item \textit{Including Tie-switches} - A typical distribution system, although radially operated, is integrated with tie-switches and planned in an open-loop configuration. In contrast to the existing literature and our prior work \cite{pesgm2017}, the proposed formulation allows modeling tie-switches and potential alternate restoration paths for the distribution system within the MILP formulation thus avoiding search-based methods.
		\item \textit{DER Reserve Energy Constraint} - The DERs or microgrids have limited reserve energy and can supply the critical loads for a fixed period of time. Our prior work \cite{pesgm2017}, did not include the model for time-dependent DER reserve energy. In this paper, the restoration time for the critical loads is optimized so that an equitable allocation of generation is obtained for each critical load.
		\item \textit{MILP formulation} - We transform the combinatorial problem of path search and power flow constraints for restoration to an MILP by appropriately defining the system variables. The alternate restoration paths due to tie-switches significantly complicate the problem. We have successfully included alternate paths and their selection into the proposed MILP formulation.
	\end{enumerate}

	\vspace{-0.4cm}

	\subsection{Assumptions}
	
		\begin{itemize} [noitemsep,topsep=0pt,leftmargin=*]
		\item Distribution circuit is equipped with enough remote-controlled switches that can be operated as per the proposed plan. The investment made under the Smart Grid Investment Grant Program managed by US Department of Energy \cite{SGIGR} has led to the deployment of remote terminal units and additional tie switches in the distribution feeders thus allowing for advanced automation capabilities \cite{alliance2014future, SGIGR, sandy, PNNL}.
		\item Non-critical loads are disconnected from the grid prior to restoring the critical loads. The recent investments made on Advanced Metering Infrastructure (AMI) has led to widespread deployment of smart meters. Smart meters allow for remote disconnect of customer supply by distribution operators \cite{smartgrid, SGI, chen2016resilient}.
		\item Radial topology of each restoration path is maintained and DERs are not networked. Although DERs can be networked during the restoration process, the operation and control of an islanded distribution network supplied by multiple DERs requires advance control function typically not available in an existing grid \cite{chen2016resilient, gao2016resilience}.
		\item  The distribution system spans relatively small geographic area and all distribution lines are equally impacted by the disaster condition with an equal probability of failure. The concept of fragility curves \cite{panteli2015grid} can be used to obtain weather-dependent failure probability for the distribution lines corresponding to a given disaster condition.
		\item The existing protection system cannot protect the restored subtree networks (RSN). With high likelihood, a suitable protection system for dynamically formed RSNs will not be available. Therefore, the faults within a RSN cannot be isolated and require de-energizing the corresponding RSN by disconnecting the respective DER.
	\end{itemize}

	\section{Restoration Problem - Definitions}
	%\vspace{-0.2cm}
	We present a graph-theoretic framework for representing distribution feeder, restored networks, and restoration objective. Each restored network is supplied by one DER and is operated in a radial topology. We aim at maximizing the resilience of the restored critical loads defined using: 1) post-restoration network reliability index and 2) DER availability index based on its lifeline performance. The problem formulation also optimizes the duration of critical load restoration. The mathematical model for the proposed problem is detailed in this section.

	\vspace{-0.4cm}
	\subsection{Graphical Representation}
	The distribution network is represented as a probabilistic graph, $G=(V,E)$, where, $V$ is the set of nodes representing buses and $E$ is the set of edges representing distribution lines. Here, a probabilistic graph is defined as a weighted graph with associated probability of success or failure for the edges. Each edge of the graph is associated with a probabilistic index, $q_e$, representing the probability of failure for the corresponding distribution line in the event of a disaster. The concept of fragility curves can be used to obtain the indices $q_e$ for a given intensity of disaster scenario \cite{panteli2015grid}. Using $q_e$, the probability of success for an edge is defined as $p_e$, where, $p_e = 1-q_e$ \cite{billinton1992reliability}.

		\begin{figure}[t]
		\centering
		\includegraphics[width=0.45\textwidth]{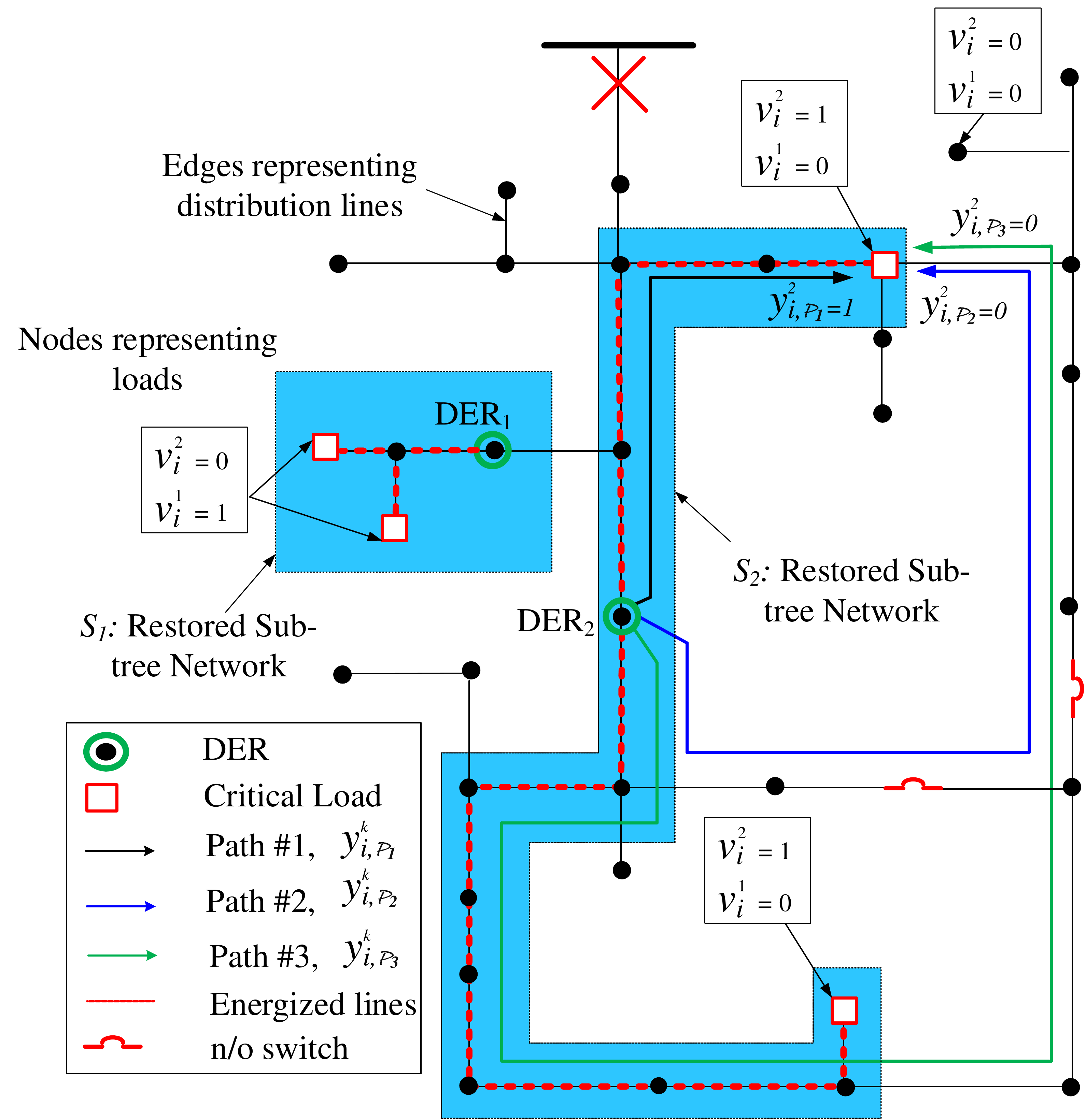}
		\caption{Example feeder and restored subtree networks (RSNs).}
		\vspace{-0.8cm}
		\label{fig:1}
	\end{figure}

The problem variables are defined as the following:
	\subsubsection{Restored Subtree Network (RSN)}
	The proposed critical load restoration approach is equivalent to decomposing the original graph $G=(V, E)$ into $n(M)$ sub-trees where $M$ is the set of nodes with DERs available for restoration. Each RSN is supplied by only one DER. Here, $S_k$ represents the restored subtree network supplied by the $k^{th}$ DER.
	
	\subsubsection{Node-DER Assignment Variable}
	Each critical load can be restored using only one DER. This is based on the assumption that DERs are not networked. A binary variable $v_i^k=\{0,1\}$ is assigned to each node, where $v_i^k=1$ implies that node $i$ is restored using DER $k$ and belongs to $S_k$, while $v_i^k=0$ implies node $i$ does not belong to $S_k$ (see Fig. \ref{fig:1}). For the distribution system with $n$ nodes and $m$ DERs, there are $m\times n$ node-DER assignment variables.

	\subsubsection{Node-Path Assignment Variable}
	Open loop configurations resulting from tie-switches will lead to multiple possible supply paths for restoring the critical loads. The restored distribution system will operate in radial configuration thus requiring a decision upon which path to use for restoration. We assign a binary variable $y_{i,\mathcal{P_\alpha}}^k=\{0,1\}$ to the nodes with multiple possible restoration paths where $y_{i,\mathcal{P_\alpha}}^k=1$ implies that node $i$ is supplied by DER $k$ following path $\alpha$ and requires energizing the set of nodes ($\mathcal{P_\alpha}$) (see Fig. \ref{fig:1}).
	
	\subsubsection{Critical Load Pickup Variable}
	The proposed approach aims at restoring all the critical loads using DERs and available feeders. However, during a natural disaster, multiple distribution lines may be at fault and some of the critical loads may not have any path available that connects them to a DER. To include the possibility of not restoring all critical loads due to system's physical constraints resulting from damages, a binary variable $s_i=\{0,1\}$ with each critical load is associated. Here, $s_i=1$ implies that critical load connected at node $i$ is picked in the restored network and vice versa.
	
\vspace{-0.4cm}	
	\subsection{Restoration Objective}
	The following three parameters characterize the restoration objective during a disaster condition: 1) total restored demand, 2) restoration duration, and 3) the post-restoration reliability of the restored subtree network (RSN). We aim at optimizing the restoration reliability (defined in later sections) and restoration duration of the system's critical loads while aiming to restore a maximum number of critical loads for a given disaster condition. The mathematical formulation for problem objective is detailed in this section. Restoration reliability depends on the: (1) probability of the RSNs remaining operational after restoration, characterized using $R_P$ and (2) the availability of the DERs. The restoration duration depends upon the energy reserve of DERs and the load profile of its restored loads.
	
	\subsubsection{Restoration Path Reliability ($R_P$)}
	Restoration path reliability characterizes the probability of the RSNs remaining operational after restoration. For DER based restoration problem, the network is said to be in operating state when there is a path from the source node (DER) to each load node restored using the DER. For $k^{th}$ DER restoring the given critical loads using $l_k$ number of distribution lines, the restoration path reliability, $R_P(k)$, is equal to the product of probability of success of each edge, $p_e$, included in the corresponding RSN, $S_k$.
	\begin{equation}\label{eq2}
	R_P(k) = p_e^{l_k}
	\end{equation}
	
	For a system with $n(M)$ DERs available for restoration, $R_P$ for the resulting restored network is given by (\ref{eq3}).
	\begin{equation}\label{eq3}
	R_P = \prod_{k=1}^{n(M)}{R_P(k)}
	\end{equation}
	
	\subsubsection{DER Availability}
	During the natural disaster, DERs are expected to operate in islanded mode. DER availability depends on lifeline performance and the configuration of the microgrid. The failure and repair rates of such configuration can be identified and used to calculate the DER availability using minimal cut-set or Markov-based methods \cite{kwasinski2012availability}. Let the availability of $k^{th}$ DER be $a_{DER}^k$. Then, its unavailability is given by $(1- a_{DER}^k)$.

	\subsubsection{Critical Load Restoration Time}
	For $k^{th}$ DER, the critical load restoration time ($T_k$) is defined as the duration for which the DER can continuously supply power to its critical loads. Let, the reserve energy of $k^{th}$ DER is $E_k$ at the time of the outage. Let, the load demand of $i^{th}$ load at time $t$ is given by $P_{i,t}$. Then, for $k^{th}$ DER, the critical load restoration time ($T_k$) is obtained by solving (\ref{eq4}). Note that $T_k$ defined in (4) is a non-linear function.
	
	\begin{equation}\label{eq4}
	\max (T_k)
	\end{equation}
	such that:
	\begin{equation}\label{}
	\bigg(E_k-\sum_{i=1}^{n(V)}v_i^k\int_{0}^{T_k}P_{i,t} dt\bigg)\geq 0 \nonumber
	\end{equation}
	
	In order to simplify the calculation for $T_k$, we assume a constant load profile for the critical loads characterized using the expected value of the actual daily load profile \cite{kwasinski2012availability}. Let, the expected demand for $i^{th}$ critical load be $P_i$. Then $\int_{0}^{T_k}P_{i,t} dt = P_{i}\times T_k$. Using  this, $T_k$ can be calculated from (\ref{eq5}).
	\begin{equation}\label{eq5}
	T_k = \dfrac{E_k}{\sum\limits_{i=1}^{n(V)}v_i^kP_{i}}
	\end{equation}
	
	In theory, each critical load should be restored for the maximum possible duration. However, maximizing the sum of restoration duration may bias the restored network such that a high capacity DER supplies a critical load with lower demand. Since restored loads are equally critical, the critical load restoration time ($T_k$) for all $S_k$ should be approximately equal while maximally utilizing the available DER capacity.
	
	All critical loads will be restored for a maximum duration when a networked system is formed where all DERs together restore all critical loads. For such a network, the critical load restoration time is given by $T_{net}$ (\ref{eq6}). For an equitable allocation of DER reserve energy to each critical load, the restoration duration for $k^{th}$ RSN, i.e. $T_k$ must be close to $T_{net}$. This constraint is defined in (6) and linearized to be incorporated as one of the network's operational constraints (see Section III.B, equation (31)).
	
	\begin{equation}\label{eq6}
	T_{net} = \dfrac{\sum\limits_{k=1}^{n(M)}E_k}{\sum\limits_{i=1}^{n(V)}P_{i}}
	\end{equation}
	
		\begin{equation}\label{eq30}
		\left|T_{net}-T_k\right| \leq \epsilon \hspace{1cm} \forall k \in M
		\end{equation}

	%\vspace{- 0.5 cm}
	\section{Resilient Restoration Problem Formulation}
	This section details the problem formulation for restoring critical loads using DERs. The objective is to maximize restoration reliability subject to feeder's operational and connectivity constraints. First, the objective function is characterized using the definitions introduced in Section II. Then several constraints are detailed for the restoration problem. A mixed-integer linear program (MILP) formulation is presented for the proposed problem detailed in (15)-(32).
	
	\subsection{Objective Function}
	The problem objective is derived from the definitions of restoration path reliability ($R_P$) and DER availability ($a^k_{DER}$). Note that $R_P$ defined in (\ref{eq3}) is a non-linear function. In this section, a linearized objective function is derived by transforming $R_P$ to a linear form.
	
	Let, $S_k$ contains $l_k$ number of distribution lines. Using the property of tree structure, the number of nodes in $S_k$ are equal to $l_k + 1$. Also, using the definition of node assignment variable, $\sum_{i = 1}^n v_i^k$ is equal to total number of nodes in $S_k$. The following condition will apply.
	\begin{equation}\label{eq7}
	\sum_{i = 1}^{n(V)} v_i^k = l_k + 1  \hspace{1 cm} \forall{k \in M }
	\end{equation}
	
	Using (1), (2), and (6), the restoration path reliability for $S_k$ and total restoration path reliability is given by (\ref{eq8}) and (\ref{eq9}), respectively.
	\begin{equation}\label{eq8}
	R_P(k) = p_e^{\sum_{i = 1}^{n(V)} v_i^k - 1}
	\end{equation}
	\begin{equation}\label{eq9}
	R_P = \prod_{k=1}^{n(M)}\left(p_e^{\sum_{i = 1}^{n(V)} v_i^k - 1}\right)
	\end{equation}
	Taking logarithm on (\ref{eq9}) with base $p_e$ results in (\ref{eq10}).
	\begin{equation}\label{eq10}
	log_{p_e}{R_P}=\sum_{k=1}^{n(M)}\bigg(\sum_{i=1}^{n(V)} (v_i^k -1) \bigg)
	\end{equation}
	
	One of the problem objectives is to maximize $R_P$. Since $p_e<1$, $log_{p_e}$ is a monotonically decreasing function, thus, transforming the problem objective as following:
	\begin{equation}\label{eq11}
	\text{maximize}(R_P) \rightarrow \text{minimize}(log_{p_e}{R_P})
	\end{equation}
	Note that, $R_P$ quantifies path reliability while $log_{p_e}{R_P}$ is a measure of probability of failure.
	
	Ignoring the constant term in (\ref{eq10}), we define a new metric, termed as effective path unavailability, $U_P$, quantifying the RSNs failure process. This term must not be confused with the existing definition of system unavailability used for continuously operated systems \cite{billinton1992reliability}. Here, $U_P$ simply provides an indication of the RSNs failing once the restoration plan has been executed.
	\begin{equation}\label{eq12}
	U_P = \sum_{k=1}^{n(M)}\bigg(\sum_{i=1}^{n(V)}v_i^k \bigg)
	\end{equation}
	
	Next, we include DER unavailability in the restoration objective defining effective restoration unavailability metric, $U_R$, in (\ref{eq13}). Here, $U_R$ is the weighted sum of $U_P$ multiplied by the unavailability of the respective DERs [16]. Once again, metric $U_R$ provides a simpler quantification for failure probability of RSNs after including the unavailability of DER. %One of the problem objectives is to minimize effective restoration unavailability ($U_R$) metric.
	\begin{equation}\label{eq13}
	\small
	U_R = \sum_{k=1}^{n(M)}\bigg((1- a_{DER}^k)\sum_{i=1}^{n(V)}v_i^k \bigg)
	\end{equation}
	
	The final restoration objective is to minimize $U_R$ while restoring a maximum number of critical loads. We define a metric, $U_{RC}$, that includes the objective of restoring a maximum number of critical loads (\ref{eq14}). Minimizing $U_{RC}$ results in the formation of a maximally reliable RSN that restores a maximum number of critical loads.
	\begin{equation}\label{eq14}
	\small
	U_{RC} = \sum_{k=1}^{n(M)}\bigg((1- a_{DER}^k)\sum_{i=1}^{n(V)}v_i^k \bigg)-n(M)n(V)\sum_{i=1}^{n(C_l)}s_i
	\end{equation}
	
	The objective function in (\ref{eq14}) is a weighted sum of two problem objectives. The first term is a metric measuring the effective restoration unavailability ($U_R$). The second term models the problem of restoring a maximum number of critical loads for a given disaster condition. The overall objective is to restore a maximum number of critical loads using available DERs by forming RSNs that are robust to post-restoration failure. The weights are assigned such that the second objective is always prioritized and the problem first ensures the restoration of a maximum number of critical loads and then minimizes the effective path unavailability.

In the proposed formulation, the post-restoration failure corresponds the second strike of the disaster events that can damage the distribution lines and cause RSNs to fail. To this regard, the formulation in (\ref{eq12}) models the probability of RSN being operational after restoration action has been executed and thus quantifies the robustness to said post-restoration failure. Recall that each RSN is a tree network or a series system comprised of a set of energized buses and distribution lines resulting in an electrical path from DER to CLs. Logically, an RSN with a lesser number of distribution lines (or nodes) will be more robust to post-restoration failure simply based on the principle of reliability for a series system \cite{billinton1992reliability}. Since the linearized formulation for effective path unavailability in (\ref{eq12}) is equal to the sum of nodes in an RSN, minimizing the term results in a robust restoration plan for post-restoration failures.
	
\vspace{-0.4cm}

\subsection{Restoration Problem Constraints}
		The several constraints associated with the proposed restoration problem are defined in this section. Let the distribution circuit be represented as a connected graph $G=(V,E)$ with the $n \in V$ number of nodes, and $l \in E$ distribution lines. Suppose, there are $m \in M$ DERs available for restoring critical loads. The constraints are described in equations (\ref{eq16})-(\ref{eq31}) and are categorized as following:

	\subsubsection{Connectivity Constraints:}
	These constraints are defined to ensure that the restored networks are connected and operate in a radial topology.
	
\begin{figure}[t]
			\centering
			\includegraphics[width=0.5\textwidth]{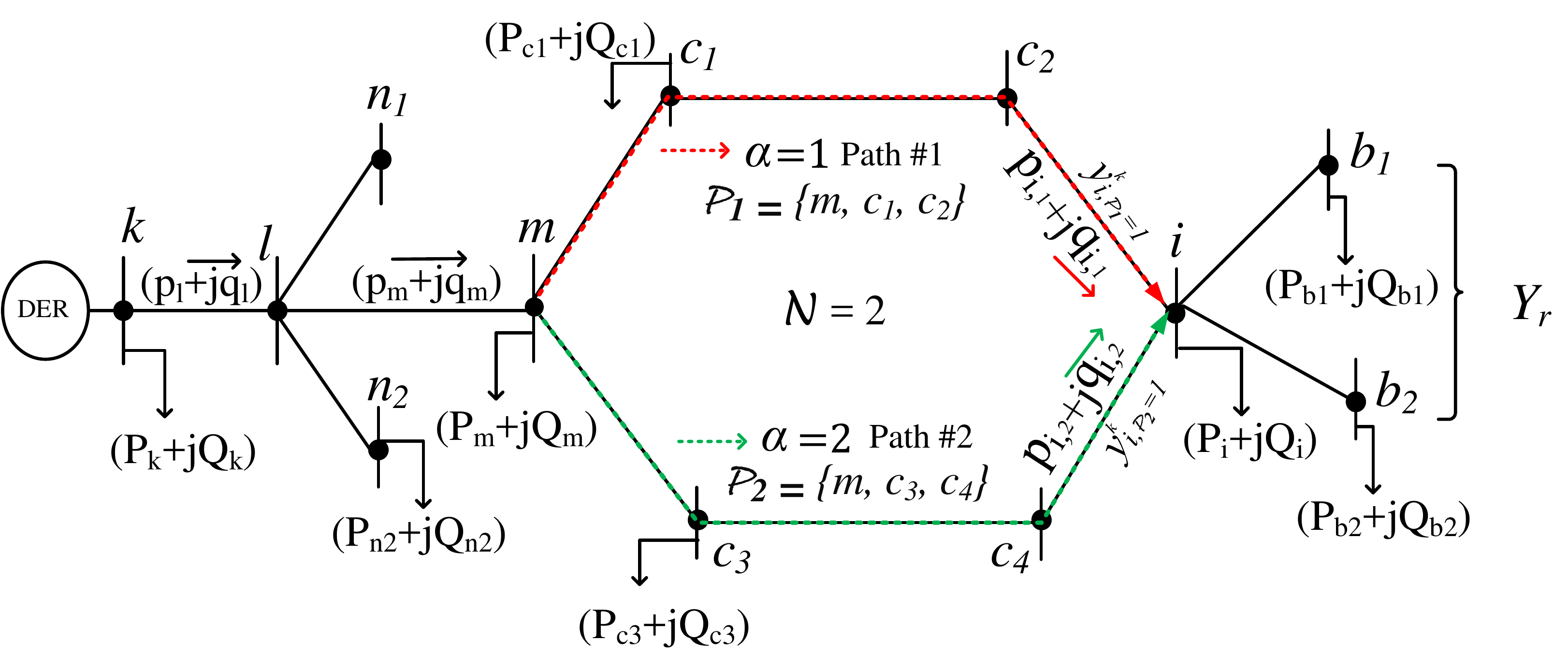}
		%	\vspace{-0.8 cm}			
			\caption{Distflow model for a distribution network operating in radial configuration with an alternate path due to a normally open tie-switch.}
			\label{fig:2}
		\vspace{-0.8cm}
		\end{figure}

	\begin{itemize}[noitemsep,topsep=0pt,leftmargin=*]
		\item Constraint (\ref{eq16}) implies that node $i$ to which DER $k$ is connected must belong to the RSN supplied by DER $k$.
		\item Constraint (\ref{eq17}) implies that the node connected to non-critical loads shall be restored at the maximum by only one RSN.
		\item Constraint (\ref{eq18}) implies that the node with a critical load must be supplied by an RSN if the critical load pickup variable $s_i=1$, and vice versa.
		\item Constraint (\ref{swi}) indicates the remote control switch unavailability. In such case, both parent and children nodes must either belong to same RSN or not belong to any RSN.
		
		\item Constraint (\ref{eq19}) characterizes the distribution line unavailability. It is possible that some distribution lines might be out of service due to the natural disaster during the restoration process. For e.g. the line joining node $i$ and $j$, $c_{ij}$, may be out of service while the corresponding nodes may still be active and even supply for critical loads. In such case, the nodes $i$ and $j$ can never belong to the same RSN.
		\item Constraint (\ref{eq20}) shall be satisfied in order to maintain the radial topology of the RSN. This equation states that a node can only belong to an RSN if its parent node belongs to the same RSN. It is important to note that the radial topology  constraint (\ref{eq20}) is only valid for the nodes that do not have alternate restoration paths. The nodes belonging to a potential loop or mesh in the distribution system can be restored using multiple restoration paths (see Fig. 1). In this case, the total number of possible paths between a source and destination is easily enumerated using Yen's k-shortest path algorithm \cite{yen1971finding}. In (\ref{eq21}), the topology constraints are defined for each available restoration paths while ensuring that only one path is energized in the restored network detailed as following.
		Let $L_p$ be the set of nodes that belong to a loop or mesh in $G$ and can be restored using a total of $\alpha$ number of paths.
		Each node in the $L_p$ is associated with node-path assignment variables ($y_{i,\mathcal{P_\alpha}}^k$) corresponding to each path $\alpha$ and DER $k$. Let, $\mathcal{P_\alpha}$ be the set of parent nodes that needs to be energized for supplying node $j$ via path-$\alpha$, and $n_{\mathcal{P_\alpha}}$ be the number of nodes in path-$\alpha$.  Then (\ref{eq21}) guarantees that all respective parent nodes are energized and the final RSN includes only one path. For example in Fig. \ref{fig:2}, if node $i$ is supplied via path-1 ($\alpha =1$), then $y_{i,\mathcal{P}_1}^k=1$ and $\mathcal{P}_1$ = \{m, $c_1$, and $c_2$\}.
	\end{itemize}

		\subsubsection{Power Flow Constraints:}
	These equations define the power flow model for the restored distribution network while incorporating alternate power flow paths and their selection in the final restored network. 	
	
	\begin{itemize}[noitemsep,topsep=0pt,leftmargin=*]
		\item Constraints (\ref{eq22}) - (\ref{eq24}) denote the branch flow and voltage constraints for a particular RSN. A linearized power flow approximation of the DistFlow \cite{yeh2012adaptive} is used in the paper. Since each RSN is a tree topology with DER at the root node, each node in the RSN has only one in-flow power. For the voltage constraints, the voltage at DER location is set to the reference value, denoted by $ V_0^k$. According to DistFlow model, the voltages at other nodes of RSN supplied by DER $k$ is given in equation (\ref{eq24}).
	\end{itemize}

%\vspace{1.5cm}

	\begin{minipage}{23.5em}
	\small
			\flushleft

		Variables: $v_i^k$, $s_i$, $p_{i,\alpha}^k$, $q_{i,\alpha}^k$, $p_i^k$, $q_i^k$, $y_{i,\mathcal{P_\alpha}}^k$, $V_{i}^k$
		\begin{flalign}\label{eq15}
		\text{Minimize:}  \ \ U_{RC}&&
		\end{flalign}
		\vspace{-0.2cm}			
		Subject to:
		\begin{flalign}\label{eq16}
		v_i^k=1, \hspace{1.3 cm} i=k, \ \ \forall k\in M&&
		\end{flalign}
		\vspace{-0.6cm}
		\begin{flalign}\label{eq17}
		\sum_{k\in M}^{} v_i^k \leq 1,\hspace{0.7 cm} \forall i\not\in C_l&&
		\end{flalign}
		\vspace{-0.4cm}
		\begin{flalign}\label{eq18}
		\sum_{k\in M}^{}v_i^k=s_i, \hspace{0.6 cm} \forall i\in C_l&&
		\end{flalign}
		\vspace{-0.4cm}
		\begin{flalign}\label{swi}
		v_i^k-v_j^k\leq1, \hspace{0.6 cm}  \forall k\in M &&
		\end{flalign}
		\vspace{-0.4cm}
		\begin{flalign}\label{eq19}
		v_i^k+v_j^k\leq1, \hspace{0.6 cm} \forall(i,j)\not\in  L_p, \ \forall k\in M &&
		\end{flalign}
		\vspace{-0.4cm}
		\begin{flalign}\label{eq20}
		v_j^k-v_i^k\leq0, \hspace{0.7 cm} \forall j\in Y,\   \forall(i,j)\not\in L_p,\  \forall k\in M &&
		\end{flalign}
		\vspace{-0.7cm}
		 
		\begin{flalign} \label{eq21}
		\begin{aligned}
		v_j^k-\frac{\sum_{i\in \mathcal{P_\alpha}}^{}v_i^k}{n_{ \mathcal{P_\alpha}}}&\leq 1-y_{j, \mathcal{P_\alpha}}^k, \hspace{0.1 cm} \ \ \forall \alpha\in \mathcal{N}\\
		\sum_{\alpha=1}^{ \mathcal{N}}1-y_{j,\mathcal{P_\alpha}}^k=& \mathcal{N}-1, \hspace{0.1 cm} \ \ \forall j\in C_l
		\end{aligned}
		&&
		\end{flalign}
		\vspace{-0.6cm}		
		
		\begin{flalign}\label{eq22}
		p_i^k=v_i^k \ P_i+\sum_{j\in Y}^{} p_j^k, \hspace{0.5 cm} \forall(i,j)\not\in L_p, \  \forall k\in M &&
		\end{flalign}
		\vspace{-0.4cm}
		\begin{flalign}\label{eq23}
		q_i^k=v_i^k \ Q_i+\sum_{j\in Y}^{} q_j^k, \hspace{0.5 cm} \forall(i,j)\not\in L_p, \  \forall k\in M &&
		\end{flalign}
		\vspace{-0.4cm}
		\begin{flalign}\label{eq24}
		V_j^k=V_i^k-\frac{r_{ij}p_j^k+x_{ij}q_j^k}{V_o^k}, \hspace{0.2 cm} \forall(i,j)\not\in L_p, \  \forall k\in M&&
		\end{flalign}
		\vspace{-0.4cm}
		\begin{flalign} \label{eq25}
		\begin{aligned}
		p_{i,\alpha}^k&=v_i^kP_i+\sum_{b\in Y_r}^{} p_b^k+\sum_{j\in Y_{\mathcal{P_\alpha}}}y_{j,\mathcal{P_\alpha}}^k\times p_{j,\alpha}^k \\
		q_{i,\alpha}^k&=v_i^kQ_i+\sum_{b\in Y_r}^{} q_b^k+\sum_{j\in Y_{\mathcal{P_\alpha}}} y_{j,\mathcal{P_\alpha}}^k\times q_{j,\alpha}^k \\ &\hspace{3.5 cm} \forall (i,j)\in L_p. \  \forall k\in M
		\end{aligned}
		&&
		\end{flalign}
		\vspace{-0.4cm}	
		\begin{flalign}\label{eq26}
		V_{j,\alpha}^k &= V_{i,\alpha}^k-\frac{r_{ij}p_{j,\alpha}^k+x_{ij}q_{j,\alpha}^k}{V_o^k}, \hspace{0.25 cm} \forall (i,j)\in L_p, \  \forall k\in M
		&&
		\end{flalign}  	
		\vspace{-0.8cm}
		
		\begin{flalign} \label{eq27}
		\begin{aligned}
		p_{i}^k&=& \sum_{\alpha=1}^{\mathcal{N}}  y_{i,\mathcal{P_\alpha}}^k \times p_{i,\alpha}^k \hspace{0.5 cm} \forall (i,j)\in L_p, \  \forall k\in M\\
		q_{i}^k&=& \sum_{\alpha=1}^{\mathcal{N}}  y_{i,\mathcal{P_\alpha}}^k\times q_{i,\alpha}^k \hspace{0.5 cm} \forall (i,j)\in L_p, \  \forall k\in M\\
		V_{j}^k&=& \sum_{\alpha=1}^{\mathcal{N}} y_{j,\mathcal{P_\alpha}}^k \times V_{j,\alpha}^k \hspace{0.5 cm} \forall (i,j)\in L_p,\  \forall k\in M
		\end{aligned}
		&&
		\end{flalign}
		\vspace{-0.6cm}		

		\begin{flalign}\label{eq28}
		\begin{aligned}
		0.95 \times v_i^k \leq V_i^k \leq 1.05 \times v_i^k, \hspace{0.5 cm} \forall i\in V, \  \forall k\in M
		\end{aligned}
		&&
		\end{flalign}
		\vspace{-0.4cm}
		\begin{flalign}\label{eq29}
		\sum_{i=1}^{n(V)} v_i^k\ P_i\leq P_{max}^k \text{  and  }
		\sum_{i=1}^{n(V)} v_i^k\ Q_i\leq Q_{max}^k&& \hspace{0.1 cm} \text{$\forall$ $k$ $\in$ $M$} 
		\end{flalign}
		\vspace{-0.3cm}
		\begin{flalign}\label{eq31}
		\frac{1}{T_{net}-\epsilon} \leq \dfrac{\sum\limits_{i=1}^{n(V)}v_i^kP_{i}}{E_k} \leq \frac{1}{T_{net}+\epsilon}  \hspace{1cm} \forall k\in M&&
		\end{flalign}
		\end{minipage}

\vspace{0.6 cm}

A node may have multiple restoration paths available due to tie-switches (see Fig. 2). In this case power flow constraints defined in (\ref{eq22}) - (\ref{eq24}) need to be modified. The new set of constraints must include the power flow along alternate restoration paths. The formulation must also ensure that only one power flow path is selected in the restored network.

\begin{itemize}[noitemsep,topsep=0pt,leftmargin=*]
	\item For nodes with multiple restoration paths, power flow equations are written along each path. The new set of power flow equations and voltage equations along each path-$\alpha$ are given in (\ref{eq25}) and (\ref{eq26}), respectively. Here, $Y_r$ is the set of children nodes of node $i$ that do not belong to the loop/mesh, $Y_{\mathcal{P_\alpha}}$ is the set of children nodes of $i$ in loop/mesh along path-$\alpha$, $p_b^k$ is the injected power of children nodes $b \in Y_r$, $p_{j,\alpha}^k$ is the injected power of children node of node $i$ from path-$\alpha$ and $y_{j,\mathcal{P_\alpha}}^k$ is the node-path assignment variable for node $j$.
	\item The path-assignment variable ($y_{i,\mathcal{P_\alpha}}^k$) is used to calculate the actual power flow values in the final restored network in (\ref{eq27}). Here, (\ref{eq27}) represents actual power flow as the sum of power flow along each path multiplied by the corresponding node-path assignment variable ($y_{i,\mathcal{P_\alpha}}^k$). Since only one path is selected in the final restored network, (\ref{eq27}) ensures actual power flow is equal to the flow along the selected path. Note that (\ref{eq25})-(\ref{eq27}) are nonlinear as they contain bilinear terms. We use Big M method to transform the bilinear constraints to a set of integer and linear constraints \cite{winston2003introduction}.
\end{itemize}

\vspace{0.3cm}
\subsubsection{Operational Constraints:}	
These constraints define the set of desired operational attributes for the restored network.
\begin{itemize}[noitemsep,topsep=0pt,leftmargin=*]
	\item Voltage at a particular node, $V_i^k$, should be within a specified range if node $i$ belongs to RSN $S_k$. Otherwise, $V_i^k$, should be zero. The logical constraints for bus voltage are eliminated as expressed in (\ref{eq28}).
\end{itemize}

\begin{itemize}[noitemsep,topsep=0pt,leftmargin=*]
	\item Constraint (\ref{eq29}) implies that the total sum of critical loads being served by DER $k$ should be less than or equal to its maximum active and reactive power capacity.
	\item Constraint (\ref{eq31}) maximizes the restoration time of each critical load while ensuring an equitable allocation of DER capacity. Here $\epsilon$ is used to represent an acceptable level of difference in restoration times of each critical load. Ideally $\epsilon$ should be equal to zero. Given different DER capacities and load demand, a small value for $\epsilon$ is specified. The restoration time constraint in (\ref{eq31}) is derived from previously defined non-linear constraint (\ref{eq30}). Since both $T_{net}$ and $\epsilon$ are constant, an equivalent linear transformation for (\ref{eq30}) is obtained in (\ref{eq31}). Note that the problem may not solve for an arbitrary value of $\epsilon$. One possible approach is to run the simulation for a number of arbitrary values of $\epsilon$, starting with the minimal possible value, and stop when a reasonable restoration plan is obtained.

\end{itemize}
	\begin{figure*}[t]
    	\centering
    	\includegraphics[width=0.85\textwidth]{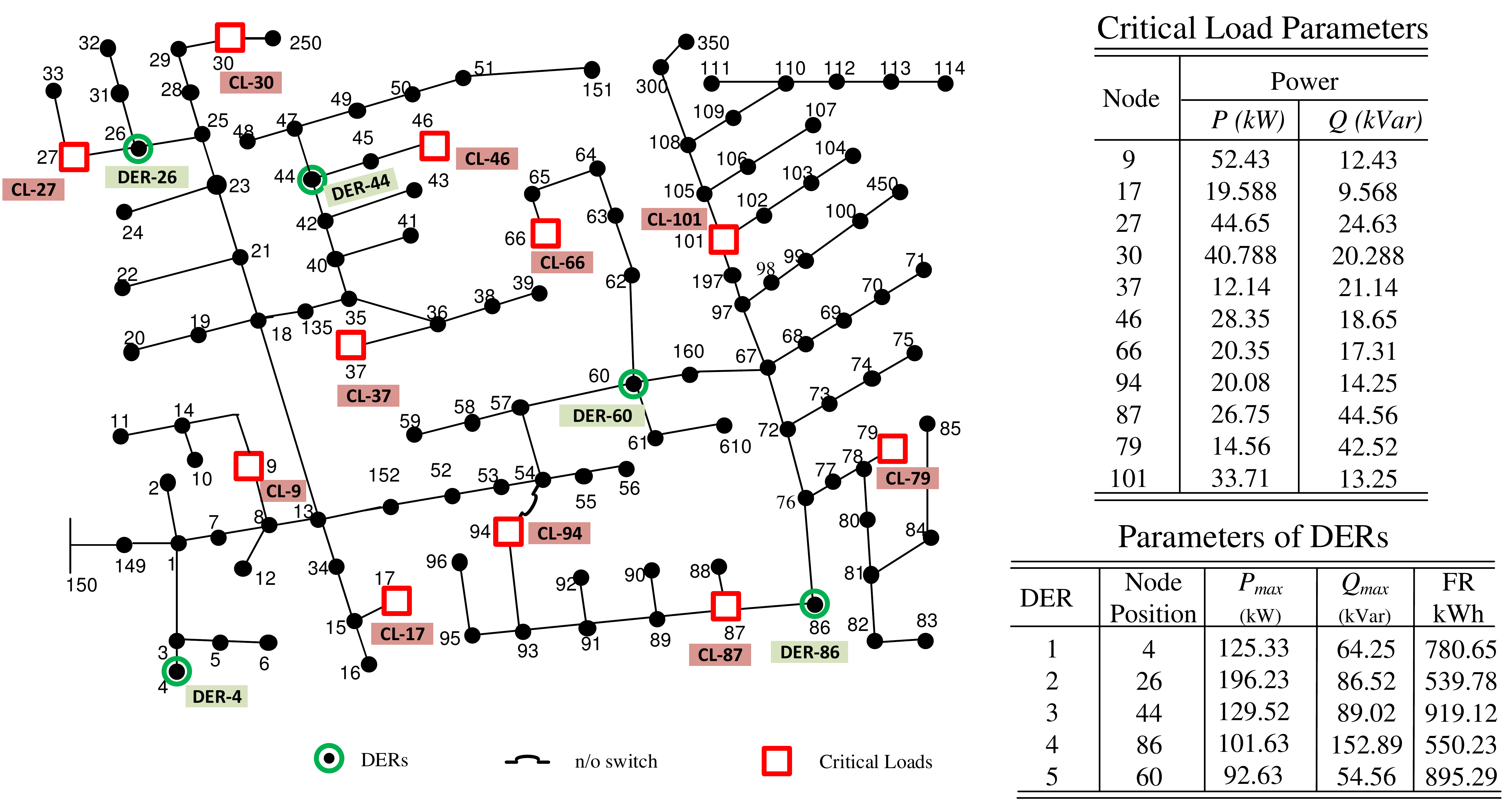}
    	\caption{IEEE 123-node distribution system with simulated locations and parameters of DERs and Critical Loads.}
    	\label{fig:3}
    	\vspace{-0.5cm}
  \end{figure*}

%\vspace{0.6cm}	
	\vspace{-0.2 cm}	
	\section{Results and Discussion}
	The proposed restoration framework is implemented on standard IEEE test systems: 123-node test feeder \cite{ieee123} and a 906-bus test feeder \cite{ieee906}. The proposed MILP formulation is solved using CPLEX 12.6. We have used MATLAB R2016a to formulate the desired model which is then linked with the CPLEX solver. The simulation is carried out on a PC with Intel Core i7-6700 @ 3.4 GHz processor and 16 GB RAM.
	
	The proposed framework is thoroughly tested using multiple case studies. For 123-node test feeder, the proposed restoration method is demonstrated for two disaster conditions: 1) disaster having a lesser impact on the distribution network resulting in only a few lines at fault, and 2) condition with a higher impact resulting in multiple faults. These cases are simulated to demonstrate that the proposed restoration approach restores a maximum number of critical loads while taking the damages in the distribution feeder into account. We also simulate cases for DERs with equal and unequal availabilities with different available capacities. Similarly, for 906-bus test feeder a case study is simulated to test scalability of the approach and applicability for a practical distribution system.

	Since the formulation discussed above approximates power flow equations, an exact power flow analysis is done for each RSN corresponding to each case study. Each RSN is simulated in OpenDSS with detailed line and load model. The actual power flow results are calculated and system losses are reported. The reported losses for each RSN are significantly small, therefore, validating the applicability of linear power flow approximation for the proposed restoration problem.
	
	Furthermore, the proposed approach is thoroughly compared with two state-of-the-art methods \cite{gao2016resilience} and \cite{chen2016resilient}. Ref. \cite{gao2016resilience} presents a method based on heuristic search. Since the approach in \cite{gao2016resilience} searches for all possible solutions, their obtained result is in fact optimal. For comparison, we show that our approach also leads to an optimal solution using mathematical optimization instead of heuristic search. Ref. \cite{chen2016resilient} uses mathematical optimization to solve a similar problem. Here, we show that our approach leads to a better solution since we include tie-switches in the formulation. In fact, for the cases when \cite{chen2016resilient} cannot restore all critical loads, our approach can restore a maximum number of critical loads by identifying alternate restoration paths with the help of tie-switches.
	
	\vspace{-0.3 cm}
	\subsection{Case Study I: IEEE 123-Node feeder System}
	
	In the simulated test case, DERs are connected to five different nodes (4, 26, 44, 60, and 86) of IEEE 123-node test system. The test system is assumed to be supplying 11 critical loads connected at nodes 9, 17, 27, 30, 37, 46, 94, 66, 101, 79, and 87. There is a normally open switch between nodes 54 and 94 that results in alternate paths. The locations of DERs and the parameters of critical loads are randomly selected for validating the proposed approach (see Fig. \ref{fig:3}). The average simulation time is 3.52 seconds.

	\vspace{-0.0cm}
	\subsubsection{Restoration during Minor Damage in Distribution Network}
	In this section, the proposed restoration strategy is tested under three different conditions considering minor damage in the distribution system due to a disaster event. It is assumed that lines 18-13 and 52-152 are at fault during the restoration process. For the remaining branches, we assume that a remote-controlled switch is available that can be operated as needed by the proposed approach to maintain the radial topology for the restored networks. The effects of DER availability and DER reserve energy on restoration unavailability and critical load restoration time are studied. As calculated from (\ref{eq6}), the maximum possible restoration time ($T_{net}$) for each critical load with a networked system together supplied by all DERs is equal to 11.75 hours. Although this is not achievable during the restoration process, an equitable allocation of generation can be obtained for each critical load by assigning a small value to $\epsilon$ (\ref{eq31}). The first two cases ignore the time constraint while in Case III, $\epsilon$ = 3 is used. It is important to note that this value shall be adjusted in case the required equitable allocation of generation cannot be obtained for each critical load.
	
	\begin{table}[t]
		\centering
		\caption{Restoration Strategy for different cases simulated for distribution system with minor damages}
		\label{singletable}
		\begin{tabular}{c|c|c|c|c}
			\toprule[0.4 mm]
			
			\multicolumn{5}{c}{    Restoration strategy for DERs with equal availability (Case I)}\\
			
			\toprule[0.4 mm]
			\hline
			\multirow{2}{*}{DERs} & Critical & Nodes on &$T_k$& Losses\\
			&Loads&Restoration Path &Hours&(\%)\\
			\hline
			\hline
			\multirow{2}{*}{DER-4}& CL-9& 4-3-1-7-8-9&\multirow{2}{*}{10.84}&\multirow{2}{*}{0.3357\%}\\
			&CL-17&4-3-1-7-8-13-34-15-17& & \\
			\hline
			\multirow{2}{*}{DER-26}& CL-27& 26-27&\multirow{2}{*}{6.31}&\multirow{2}{*}{0.0277\%}\\
			&CL-30&26-25-28-29-30& & \\
			\hline
			\multirow{2}{*}{DER-44}& CL-37& 44-42-40-35-36-37&\multirow{2}{*}{22.58}&\multirow{2}{*}{0.2154\%}\\
			&CL-46&44-45-46& & \\
			\hline
			\multirow{3}{*}{DER-86}& CL-79& 86-76-77-78-79&\multirow{3}{*}{7.33}&\multirow{3}{*}{0.1109\%}\\
			&CL-87&86-87& & \\
			&CL-101&86-76-72-67-97-197-101& & \\
			\hline
			\multirow{2}{*}{DER-60}& CL-66& 60-62-63-64-65-66&\multirow{2}{*}{22.144}&\multirow{2}{*}{0.2608\%}\\
			&CL-94&60-57-54-94& &\\
			\toprule[0.4 mm]

			\multicolumn{5}{c}{    Restoration strategy for DERs with unequal availability (Case II)}\\
			\toprule[0.4 mm]
			\hline
			\multirow{2}{*}{DERs} & Critical & Nodes on &$T_k$& Losses\\
			&Loads&Restoration Path &Hours& (\%)\\
			\hline
			\hline
			\multirow{2}{*}{DER-4}& CL-9& 4-3-1-7-8-9& \multirow{2}{*}{10.84}&\multirow{2}{*}{0.3357\%}\\
			&CL-17&4-3-1-7-8-13-34-15-17& & \\
			\hline
			\multirow{4}{*}{DER-26}& CL-27& 26-27& \multirow{4}{*}{5.53}&\multirow{4}{*}{0.0789\%}\\
			&CL-30&26-25-28-29-30& & \\
			&\multirow{2}{*}{CL-37}&26-25-23-21-18-& & \\
			&&135-35-36-37& & \\
			\hline
			\multirow{1}{*}{DER-44}& CL-46& 44-45-46& \multirow{1}{*}{32.18}&\multirow{1}{*}{0.0996\%}\\
			\hline
			\multirow{4}{*}{DER-86}& CL-79& 86-76-77-78-79& \multirow{4}{*}{5.78}&\multirow{4}{*}{0.1490\%}\\
			&CL-87&86-87& & \\
			&CL-94&86-87-89-91-93-94& & \\
			&CL-101&86-76-72-67-97-197-101& & \\
			\hline
			\multirow{1}{*}{DER-60}& CL-66& 60-62-63-64-65-66& \multirow{1}{*}{43.99}&\multirow{1}{*}{0.1741\%}\\
			\toprule[0.4 mm]
			
			\multicolumn{5}{c}{    Restoration strategy for DERs with unequal availability}\\
			\multicolumn{5}{c}{    and time constraints included  (Case III)}\\
			\toprule[0.4 mm]
			\hline
			\multirow{2}{*}{DERs} & Critical & Nodes on & $T_k$& Losses\\
			&Loads&Restoration Path &Hours&(\%)\\
			\hline
			\hline
			\multirow{2}{*}{DER-4}& CL-9& 4-3-1-7-8-9&\multirow{2}{*}{10.84}&\multirow{2}{*}{0.3357\%}\\
			&CL-17&4-3-1-7-8-13-34-15-17& & \\
			\hline
			\multirow{1}{*}{DER-26}& CL-27& 26-27&\multirow{1}{*}{12.08}&\multirow{1}{*}{0.0131\%}\\
			\hline
			\multirow{4}{*}{DER-44}& CL-46& 44-45-46& \multirow{4}{*}{11.28}&\multirow{4}{*}{0.1598\%}\\
			&CL-37&44-42-40-35-36-37& & \\
			&\multirow{2}{*}{ CL-30}&44-42-40-35-135-18-& & \\
			&&21-23-25-28-29-30& & \\
			\hline
			\multirow{2}{*}{DER-86}& CL-79& 86-76-77-78-79& \multirow{2}{*}{13.32}&\multirow{2}{*}{0.1668\%}\\
			&CL-87&86-87& & \\
			\hline
			\multirow{2}{*}{DER-60}& CL-101& 60-160-67-97-197-101&\multirow{2}{*}{12.07}&\multirow{2}{*}{0.4627\%}\\
			&CL-66&60-62-63-64-65-66& & \\
			&CL-94&60-57-54-94& & \\
			\toprule[0.4 mm]
			\hline
		\end{tabular}
		\vspace{-0.2 cm}
	\end{table}

	\paragraph{Case I - DERs with Equal Availability}
	In this case, we assume that each DER available for restoration has an equal availability, $a_{DER}^k = 0.95$. The objective function in this case transforms to simply minimizing the total number of nodes present in the restored subtree networks. This is because $U_R$ is weighted with same unavailability $(1-a^k_{DER})$ for all DERs. The results for optimal restoration plan are shown in Table \ref{singletable} where five restored subtree networks are formed, each energized by one DER. The restoration path for each critical load is also reported in Table I.
	
	\paragraph{Case II - DERs with Unequal Availability}
	In this case, we assume that DERs have unequal availability. The DERs located at nodes 44 and 60 have an availability equal to 0.92 and 0.90, respectively, less than that of Case I. Unlike the previous case, the objective function in this case is to minimize the total number of nodes present in the restored subtree networks weighted by the respective unavailability of the DERs. A different restoration topology is obtained for critical loads located at nodes 37 and 94. This is because the DERs at nodes 44 and 60 have less availability than before and our aim is to minimize the restoration unavailability. The restoration path for each critical load is given in Table \ref{singletable}.

	\paragraph{Case III - With Critical Load Restoration Time Constraint}
	In addition to the unequal availability of DERs in Case II, this case takes into account the time constraints described by equation (\ref{eq31}).
	Since the previous two cases are focused on minimizing the restoration unavailability only, the sharing of energy among DERs is ignored. A particular scenario is then observed when a DER with high fuel reserve is supplying the critical load with lower demand or vice-versa leading to an unacceptable level of bias. For example, in Case II, RSN formed with DER 60 has restoration time of 43.99 hours while RSN formed with DER 86 has restoration time of 5.78 hours only. Since all critical loads are equally important, we impose the constraint defined in equation (\ref{eq31}) in order to remove bias such that DERs can share their energy and still maintain the minimum possible unavailability. The restoration path for each critical load and restoration time is given in Table \ref{singletable}. It is observed that all critical loads in the distribution network are now picked up for a time close to $T_{net} = 11.75$ hours.
\begin{table}[t]
\vspace{-0.3cm}
	\centering
	\caption{Summary of the case studies for distribution system with minor damages}
\vspace{-0.3cm}
	\label{discuss}
	\begin{tabular}{c|c|c}
		\hline
		\hline
		Case Studies & Average Bias (Hours) &  $U_R$\\
		\hline
		Case I& 2.09& 2.2\\
		Case II& 7.714&2.74\\
		Case III& 0.168&3.58\\
		\hline
		\hline
	\end{tabular}
	\vspace{-1.5 cm}	
\end{table}

	An equal critical load restoration time, however, comes at the cost of increased effective unavailability index $U_R$. From Table \ref{discuss}, cases I and II have a relatively smaller value of effective restoration unavailability as compared to Case III. This is because in Case III, instead of just minimizing the number of nodes, the algorithm also needs to satisfy an equitable allocation of DER capacities which results in longer restoration paths for a few RSNs. After including the time constraint, the average bias among the RSN serving critical loads is significantly reduced (see Table \ref{discuss}). The average bias for each case is calculated by averaging the difference between $T_{net}$ and $T_k$ for each RSN.

	\subsubsection{Restoration during Major Damage in Distribution Network}
	In this section, the proposed restoration strategy is tested for the case when distribution system sustains multiple faults (Lines 26-27, 13-18, 51-151, 91-93, 54-57, and 67-97) due to the disaster condition. The DERs located at nodes 44 and 86 have availability of 0.92 and 0.90, respectively and rest of the DERs have availability of 0.95. In this case, with major disruptions in distribution network, the priority is to restore a maximum number of critical loads disregarding the equitable allocation of DER capacities. Hence, the objective is to restore the maximum possible critical loads by forming restored networks of minimum unavailability.
	The results are shown in Table \ref{table:6}. Because of multiple line faults, CL-27 and CL-101 remain unserved in the restoration process as there is no path available for supplying these loads (see Fig. \ref{fig:3}). The restoration path for the remaining 9 critical loads and the restoration unavailability of the respective restored subtree network are detailed in Table \ref{table:6}.	

	\begin{table}[t]
		\centering
		\caption{ Restoration strategy for the distribution system sustaining a major damage with multiple faults}
		\label{table:6}
		\begin{tabular}{c|c|c|c|c}
			\hline
			\hline
			\multirow{2}{*}{DERs} & Critical & Nodes on &\multirow{2}{*}{$U_R^k$}& Losses\\
			&Loads&Restoration Path && (\%)\\
			\hline
			\hline
			\multirow{4}{*}{DER-4}& CL-9& 4-3-1-7-8-9& \multirow{4}{*}{0.75}&\multirow{4}{*}{0.425\%}\\
			&CL-17&4-3-1-7-8-13-34-15-17& & \\
			&\multirow{2}{*}{CL-94}&4-3-1-7-8-13-152-& & \\
			&&52-53-54-94& & \\
			\hline
			\multirow{3}{*}{DER-26}& CL-30& 26-25-28-29-30& \multirow{3}{*}{0.6}&\multirow{3}{*}{0.1789\%}\\
			&\multirow{2}{*}{CL-37}&26-25-23-21-18-& & \\
			&&135-35-36-37& & \\
			\hline
			\multirow{1}{*}{DER-44}& CL-46& 44-45-46& \multirow{1}{*}{0.24}&\multirow{1}{*}{0.0996\%}\\
			\hline
			\multirow{1}{*}{DER-86}& CL-87& 86-87& \multirow{1}{*}{0.2}&\multirow{1}{*}{0.0409\%}\\
			\hline
			\multirow{2}{*}{DER-60}& CL-66& 60-62-63-64-65-66& \multirow{2}{*}{0.65}&\multirow{2}{*}{0.1941\%}\\
			&CL-79&60-160-67-72-76-77-78-79& & \\
			\hline
			\hline
		\end{tabular}
	\vspace{-1.8 cm}
	\end{table}

\vspace{-0.2 cm}	

	\begin{figure}[b]
		\centering
		\includegraphics[width=0.47\textwidth]{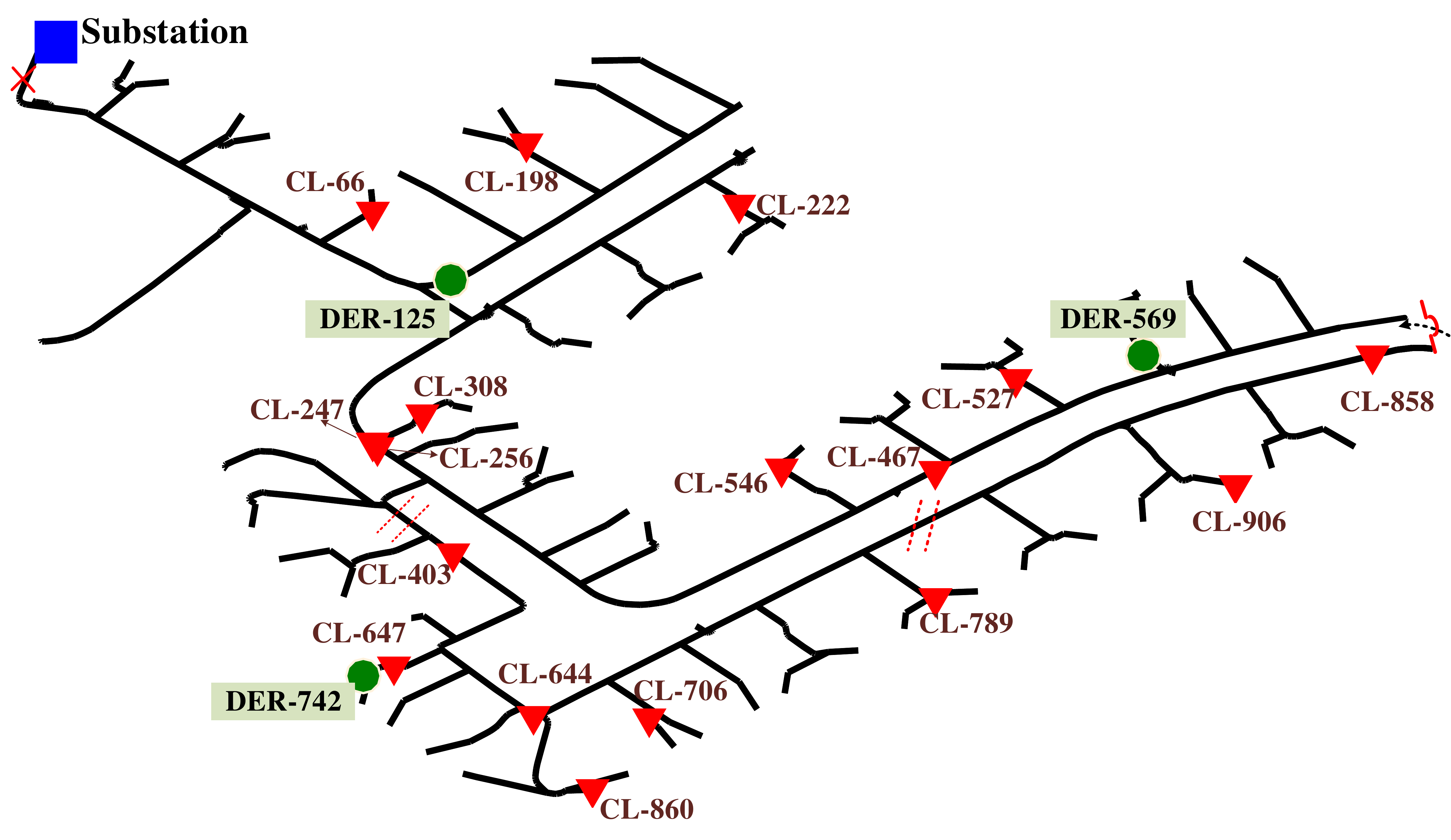}
		\caption{IEEE 906-bus test feeder with simulated locations of DERs and Critical Loads.}
	%	\vspace{-3cm}
		\label{fig:4}
	\end{figure}

\subsection{Case Study II: IEEE 906-bus Low-Voltage Test Feeder}
In this low-voltage distribution feeder, there are 906 nodes where the main feeder and laterals are at voltage level of 416 V. It is assumed that DERs are connected at three different nodes (125, 568, and 742) and the system is supplying 17 critical loads. The parameters of DERs and critical loads are shown in Table \ref{parder} and Table \ref{parcl} respectively.

The locations of DERs and the parameters of critical loads are randomly selected in order to validate the proposed approach. It is assumed that, after a disaster, the line switches 378-384 and 762-770 are in open status or at fault and the feeder is disconnected from the main supply (see Fig. \ref{fig:4}). There is a normally open switch between nodes 618 and 881 that results in loop configuration.

In this case, we assume that each DER available for restoration has an equal availability, $a_{DER}^k = 0.95$. The results for optimal restoration plan are shown in Table \ref{realtable} where three restored subtree networks are formed, each energized by one DER. Because of the line faults and DER capacity constraints, CL-860 is not supplied in the restoration process. Please note that nodes on restoration path are not reported in the table because of the space constraint. The average simulation time is 7.69 seconds.

	It should be noted that this paper presents a generic framework for critical load restoration during a disaster condition. The parameters and network model used in the formulation can be easily replaced to represent a real-world system and a real-world operational scenario. The results will differ for different systems with varying levels of DER penetrations and automation capabilities. The formulation, however, is always applicable and will result in a feasible restoration plan.

	\begin{table}[htbp]
	\centering
	\caption{Parameters of DERs for IEEE 906-bus test feeder}
	\label{parder}
	\begin{tabular}{ccccc}
		\toprule
		\multirow{2}{*}{DER} & Node & $P_{max}$ &$Q_{max}$ & Availability
		\\
		%\cline{5-6}
		&Position&kW&kVar&\\
		\hline
		1& 125&442.65&170.25&0.95\\
		2&569&321.52&93.25&0.95\\
		3&742&309.40&112.58&0.95\\
		\toprule
	\end{tabular}
	\vspace{-10 pt}
\end{table}

\begin{table}[t]
	\centering
	\caption{Parameters of Critical Loads for IEEE 906-bus test feeder}
	\label{parcl}
	\begin{tabular}{cccccc}
		\toprule
		Node & P (kW) &Q (kVar) & Node & P (kW) &Q (kVar) \\
		%\cline{5-6}
		\hline
		66& 56.44&12.53&546&48.40&12.32\\
		198& 93.85&19.80&644&99.18&9.56\\
		222& 30.40&22.52&647&61.13&26.63\\
		247& 84.58&42.37&706&30.95&6.52\\
		256& 50.84&28.56&789&71.56&34.89\\
		308& 98.23&19.20&860&16.30&2.36\\
		403& 33.17&5.26&858&20.99&10.23\\
		467& 41.81&13.24&906&31.98&5.63\\
		527& 85.83&28.63&-&-&-\\
		\toprule
	\end{tabular}
	\vspace{-0.2 cm}
\end{table}	

\begin{table}[!ht]
		\centering
		\caption{ Restoration strategy for the 906-bus IEEE Test System}
		\label{realtable}
		\begin{tabular}{c|c|c|c}
			\hline
			\hline
			DERs & Critical Loads Picked Up  &$U_R^k$&Losses(\%)\\
			\hline
			\hline
			DER-125& 66, 198, 222, 247, 256, \& 308 &3.7&0.0285\%\\
			\hline
			DER-569& 467, 546, 527, 858, \& 906 &3.65&0.011\%\\
			\hline
			DER-742& 403, 644, 647, 706, \& 789 &4.15&0.074\%\\
			\hline
			\hline
		\end{tabular}
		\vspace{-6 pt}
	\end{table}

	\vspace{-0.3 cm}
	
	\subsection{Performance Comparison}	
	In this section, the proposed approach is thoroughly compared with two state-of-art methods \cite{gao2016resilience} and \cite{chen2016resilient}. The approach presented in \cite{gao2016resilience} is based on heuristic search method while \cite{chen2016resilient} presents an MILP problem formulation for critical load restoration problem. Test cases are simulated using IEEE test feeders and different algorithms are compared for their ability to restore the critical loads.

	\subsubsection{Comparison with Heuristic Method in \cite{gao2016resilience}}
	
	A two-stage approach based on search-based method is proposed for the critical load restoration problem in \cite{gao2016resilience}. We compare our approach with the one proposed in \cite{gao2016resilience}. The circuit topology and parameters are chosen same as that of Case Study I-1 i.e., IEEE 123-node test system for the scenario with minor damages to the distribution system. As proposed in \cite{gao2016resilience}, the restoration path for each possible DER-CLs combinations are searched and saved in a strategy table. The feasible restoration paths are then obtained to form an updated strategy table. The original optimization problem presented in \cite{gao2016resilience} aims at maximizing the weighted sum of prioritized critical loads. Since, our formulation assumes all CLs to be equally critical, for a reasonable comparison, the objective function is defined with equal weights for all CLs. Finally, using the approach presented in \cite{gao2016resilience}, a restoration plan is obtained by selecting feasible paths that minimize the objective function.
		
	For the selected test system, the strategy table at first stage consists of $16,431$ restoration paths. Note that the test case is comprised of $5$ DERs supplying $11$ critical loads. Restoration paths containing the faulted lines and networked DERs are removed such that the updated strategy table consists of only $54$ feasible restoration paths. Table \ref{compare2} shows the restoration strategy which is obtained by selecting optimal paths out of all feasible paths. The average simulation time for storing and updating the strategy table is 13.8 minutes and for optimization is 1.23 seconds. Recall that the average simulation time for the approach proposed in this paper is 3.52 seconds. The results obtained using the method proposed in this paper for the same case study and for same objective function is shown in Table I Case III. It should noted that both \cite{gao2016resilience} and the approach presented in this paper result in same restoration plan. This is because the method proposed in \cite{gao2016resilience} is based on exhaustive search and will always result in an optimal plan since its search space is comprised of all possible restoration options. On the contrary, our approach obtains an optimal solution using mathematical optimization techniques. Furthermore, unlike \cite{gao2016resilience}, the proposed approach does not require to store a strategy table since it dynamically obtains optimal restoration plan after a disaster. %Here, the topological constraints for loop configuration are defined for each available restoration paths while ensuring that only one path is selected in the final restoration process. Power flow equations are modified as given in equations (26) \& (27) and written along each possible path. Finally, a suitable path minimizing the objective function is obtained using path assignment variable.

A search-based formulation is usually employed because it is difficult to obtain closed-form expressions for path selection and power flow equations for a system with multiple restoration paths. The approach presented in this paper models topological and power flow constraints for multiple restoration paths. A binary variable is defined to model path selection problem along with restoration as an MILP. The search-based methods are, therefore, avoided.

	\begin{table}[!t]
	\centering
	\caption{ Restoration strategy using heuristic method \cite{gao2016resilience} for distribution system with minor damages}
	\label{compare2}
	\begin{tabular}{c|c|c|c|c}
		\hline
		\hline
		\multirow{2}{*}{DERs} & Critical & Nodes on & $T_k$& Losses\\
		&Loads&Restoration Path &Hours&(\%)\\
		\hline
		\hline
		\multirow{2}{*}{DER-4}& CL-9& 4-3-1-7-8-9&\multirow{2}{*}{10.84}&\multirow{2}{*}{0.3357\%}\\
		&CL-17&4-3-1-7-8-13-34-15-17& & \\
		\hline
		\multirow{1}{*}{DER-26}& CL-27& 26-27&\multirow{1}{*}{12.08}&\multirow{1}{*}{0.0131\%}\\
		\hline
		\multirow{4}{*}{DER-44}& CL-46& 44-45-46& \multirow{4}{*}{11.28}&\multirow{4}{*}{0.1598\%}\\
		&CL-37&44-42-40-35-36-37& & \\
		&\multirow{2}{*}{ CL-30}&44-42-40-35-135-18-& & \\
		&&21-23-25-28-29-30& & \\
		\hline
		\multirow{2}{*}{DER-86}& CL-79& 86-76-77-78-79& \multirow{2}{*}{13.32}&\multirow{2}{*}{0.1668\%}\\
		&CL-87&86-87& & \\
		\hline
		\multirow{2}{*}{DER-60}& CL-101& 60-160-67-97-197-101&\multirow{2}{*}{12.07}&\multirow{2}{*}{0.4627\%}\\
		&CL-66&60-62-63-64-65-66& & \\
		&CL-94&60-57-54-94& & \\
		\hline
		\hline
	\end{tabular}
	\vspace{-1cm}
\end{table}	 		
It should be noted that, \cite{gao2016resilience} may result in multiple solutions that can restore a maximum number of critical loads. One of the solution will consist of RSNs with a minimum number of nodes leading to a robust restoration strategy. The original formulation proposed in \cite{gao2016resilience} does not consider minimizing the number energized nodes in the formulation. We have included this condition when implementing \cite{gao2016resilience}. This is the reason the results obtained for \cite{gao2016resilience} and our method are the same.

	\subsubsection{Comparison with MILP Formulation in \cite{chen2016resilient}}
	The proposed approach is also compared with the MILP formulation presented in \cite{chen2016resilient}. It should be noted that the approach proposed in \cite{chen2016resilient} does not include tie-switches in the restoration problem formulation. Tie-switches could be very beneficial in restoring critical loads when distribution system suffers from multiple damages. The approach proposed in this paper incorporates tie-switch in the distribution network and provides flexibility for path selection during the restoration process. A comparison is made using both IEEE 123-node and IEEE 906-bus distribution feeder. The circuit topology and parameters of IEEE 123-node are chosen same as that of Case Study I-2 i.e., the case with major damages to the distribution feeder. For 906-bus system, circuit topology and parameters are similar to that of Case Study II.

	\begin{table}[t]
		\centering
\vspace{-0.3cm}
		\caption{ Performance Comparison of proposed approach and \cite{chen2016resilient} for modified IEEE-123 node test case.}
		\label{compare1}
		\begin{tabular}{c|c|c|c|c}
			\hline
			\hline
			\multirow{2}{*}{DERs} & \multicolumn{2}{c|}{Proposed Approach}&\multicolumn{2}{c}{\cite{chen2016resilient}} \\
			\cline{2-5}
			& CLs Picked&$U_R^k$&CLs Picked&$U_R^k$\\
			\hline
			\hline
			DER-4&9, 17, 94 &0.75&9, 17 &0.5\\
			\hline
			DER-26& 30, 37 &0.6& 30, 37 &0.6\\
			\hline
			DER-44& 46 &0.24& 46 &0.24\\
			\hline
			DER-86& 87 &0.2&87 &0.2\\
			\hline
			DER-60& 66, 79 &0.65& 66, 79 &0.65\\
			\hline
			%&$U_{RC}$&-897.56&$U_{RC}$&-797.81\\	
			%\hline
			\hline
		\end{tabular}
		\vspace{-0.1cm}
	\end{table}
	
	The restoration plans for 123-node system and 906-bus distribution feeder is shown in Table \ref{compare1} and Table \ref{compare3}, respectively. It should be noted that for both test feeders, our approach helps to restore more critical loads than the one using \cite{chen2016resilient}. For example, for 123-node feeder, CL-94 is not picked by the approach presented in \cite{chen2016resilient}. Similarly, for 906-bus feeder, CL-858 and CL-906 are not restored when using \cite{chen2016resilient}.  %It should be observed that with proposed approach, the effective restoration unavailability of RSNs ($U_{RC}$) is smaller than that of method proposed in \cite{chen2016resilient}.
Therefore, it is shown that, when distribution system is sustaining a major damage with multiple faults, tie-switches may help in restoring additional critical loads. The presented approach in this paper results in a restoration plan that restores a maximum number of critical loads. Since \cite{chen2016resilient} does not include the formulation for alternate restoration path due to tie-switches, some of the critical loads cannot be restored in case a disaster leads to major damages to the distribution network.

	\begin{table}[t]
		\centering
		\caption{Performance Comparison of proposed approach and \cite{chen2016resilient} for IEEE 906-bus test feeder.}
		\label{compare3}
		\begin{tabular}{c|c|c|c|c}
			\hline
			\hline
			\multirow{2}{*}{DERs} & \multicolumn{2}{c|}{Proposed Approach}&\multicolumn{2}{c}{\cite{chen2016resilient}} \\
			\cline{2-5}
			&CLs Picked&$U_R^k$&CLs Picked&$U_R^k$\\
			\hline
			\hline
			\multirow{2}{*}{DER-125} & 66, 198, 222, &\multirow{2}{*}{3.7}&66, 198, 222&\multirow{2}{*}{3.7}\\
			&247, 256, 308 &&247, 256, 308 &\\
			\hline
			\multirow{2}{*}{DER-569} & 467, 546, 527 &\multirow{2}{*}{3.65}&\multirow{2}{*}{467, 546, 527}&\multirow{2}{*}{1.4}\\
			&858, 906 &&&\\
			\hline
			\multirow{2}{*}{DER-742} & 403, 644, 647 &\multirow{2}{*}{4.15}&403, 644, 647&\multirow{2}{*}{4.15}\\
			&706, 789 &&706, 789 &\\
			\hline
			%&$U_{RC}$&-43,488&$U_{RC}$&-38,052\\	
			%\hline
			\hline
		\end{tabular}
	\vspace{-0.9 cm}
	\end{table}

	\section{Conclusions}
	In this paper, we proposed a novel method to restore critical loads in distribution circuit when the main grid is not available as a result of a natural disaster. The proposed framework is generic and the parameters and network models used in the formulation can be easily tweaked to represent a real-world system and a real-world fault/damage scenario. A new metric is defined to quantify the restoration unavailability ($U_R$) for the restored distribution circuit and an MILP problem is formulated to obtain a robust restoration plan while satisfying operational and connectivity constraints. Open-loop configuration of the distribution network is included in the optimization problem and a suitable path is selected in the restoration plan while maintaining a radial operation. Critical load restoration time constraint is also incorporated to ensure an equitable allocation of generation resources to each critical load. The simulation results show that the restored topology is able to take both network failure probability and DER availability into account in coming up with a restoration plan. In addition, it is also demonstrated that the proposed approach restores a maximum number of critical loads while taking into account the damages within distribution feeder. By minimizing the effective unavailability metric for the restored circuit, a restoration plan is envisioned that is robust to the post-restoration failures resulting from a second strike of disaster event potentially damaging the distribution lines.

	\bibliographystyle{IEEEtran}
	%\vspace{-0.1cm}
	\bibliography{references}	
	\vspace{-1.0 cm}
    
    \begin{IEEEbiography}[{\includegraphics[width=1in,height=1.25in,clip,keepaspectratio]{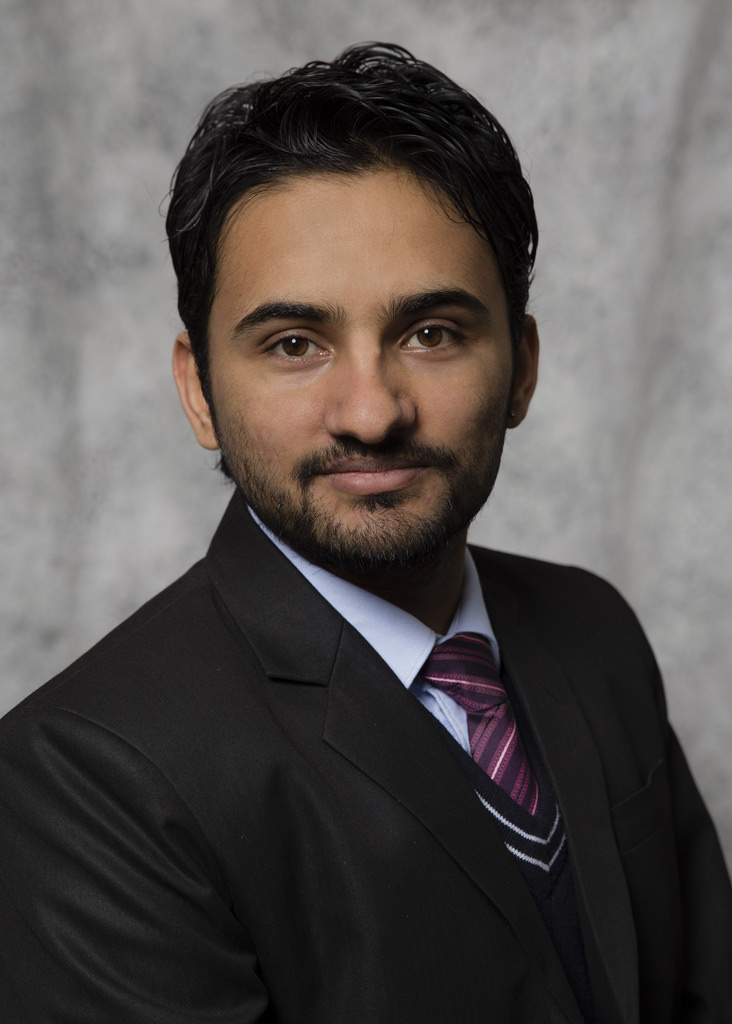}}]{\textbf{Shiva Poudel}} (S'15) received the B.E. degree from the Department of Electrical Engineering, Pulchowk Campus, Kathmandu, Nepal, in 2013, and the M.S. degree from the Electrical Engineering and Computer Science Department, South Dakota State University, Brookings, SD, USA, in 2016. He is now pursuing the Ph.D. degree in the School of Electrical Engineering and Computer Science, Washington State University, Pullman, WA. 
	 His current research interests include distribution system restoration, resilience assessment, and distributed algorithms.
		\vspace{-0.5 cm}
	\end{IEEEbiography}
	
	\begin{IEEEbiography}[{\includegraphics[width=1in,height=1.25in,clip,keepaspectratio]{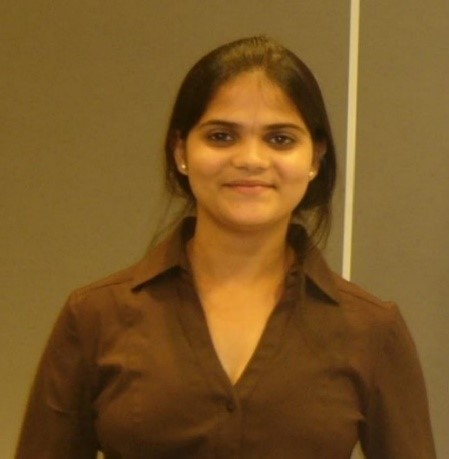}}]{\textbf{Anamika Dubey}} (M'16) received the M.S.E and Ph.D. degrees in Electrical and Computer Engineering from the University of Texas at Austin in 2012 and 2015, respectively. Currently, she is an Assistant Professor in the School of Electrical Engineering and Computer Science at Washington State University, Pullman.
	Her research focus is on the analysis, operation, and planning of the modern power distribution systems for enhanced service quality and grid resilience. At WSU, her lab focuses on developing new planning and operational tools for the current and future power distribution systems that help in effective integration of distributed energy resources and responsive loads. 
		%\vspace{-2 cm}
		\vspace{-1 cm}
	\end{IEEEbiography}

\end{document}